\documentclass[journal]{IEEEtran}
\usepackage{cite}
\usepackage{amsmath,amssymb,amsfonts}
\usepackage{algorithm} 
\usepackage{subfigure}
\usepackage{textcomp}
\usepackage{bm}
\usepackage{amsmath,amsfonts}
\usepackage{algorithmic}
\usepackage{array}
\usepackage[caption=false,font=normalsize,labelfont=sf,textfont=sf]{subfig}
\usepackage{textcomp}
\usepackage{stfloats}
\usepackage{url}
\usepackage{verbatim}
\usepackage{graphicx}
\usepackage{color}
\def\BibTeX{{\rm B\kern-.05em{\sc i\kern-.025em b}\kern-.08em
    T\kern-.1667em\lower.7ex\hbox{E}\kern-.125emX}}
\usepackage{balance}
\begin{document}

\title{Adaptive Digital Twin for UAV-Assisted Integrated Sensing, Communication, and Computation Networks}

\author{Bin Li, \IEEEmembership{Member, IEEE}, 
	Wenshuai Liu, Wancheng Xie, 
	Ning Zhang \IEEEmembership{Senior Member, IEEE}, \\
	and Yan Zhang, \IEEEmembership{Fellow, IEEE}
	
\thanks{This work was supported in part by the National Natural Science Foundation of China under Grant 62101277, in part by the Natural Science Foundation of Jiangsu Province under Grant BK20200822, and in part by the open research fund of Key Lab of Broadband Wireless Communication and Sensor Network Technology (Nanjing University of Posts and Telecommunications), Ministry of Education under Grant JZNY202103. (\textit{Corresponding author: Bin Li.})}
	
\thanks{B. Li is with the School of Computer Science, Nanjing University of Information Science and Technology, Nanjing 210044, China, and also with the Key Lab of Broadband Wireless Communication and Sensor Network Technology (Nanjing University of Posts and Telecommunications), Ministry of Education, Nanjing 210003, China (e-mail: bin.li@nuist.edu.cn).}
\thanks{W. Liu and W. Xie are with the School of Computer Science, Nanjing University of Information Science and Technology, Nanjing 210044, China (e-mail: liuwenshuai@nuist.edu.cn; zuoyeyiwancheng@gmail.com).}
\thanks{N. Zhang is with the Department of Electrical and Computer Engineering, University of Windsor, Windsor, ON N9B 3P4, Canada (e-mail: ning.zhang@uwindsor.ca).}
\thanks{Y. Zhang is with University of Oslo, Norway, and also with Simula Metropolitan Center for Digital Engineering, Norway (e-mail: yanzhang@ieee.org).}
}

\maketitle

\begin{abstract}
In this paper, we study a digital twin (DT)-empowered integrated sensing, communication, and computation network. Specifically, the users perform radar sensing and computation offloading on the same spectrum, while unmanned aerial vehicles (UAVs) are deployed to provide edge computing service. We first formulate a multi-objective optimization problem to minimize the beampattern performance of multi-input multi-output (MIMO) radars and the computation offloading energy consumption simultaneously. Then, we explore the prediction capability of DT to provide intelligent offloading decision, where the DT estimation deviation is considered. To track this challenge, we reformulate the original problem as a multi-agent Markov decision process and design a multi-agent proximal policy optimization (MAPPO) framework to achieve a flexible learning policy. Furthermore, the Beta-policy and attention mechanism are used to improve the training performance. Numerical results show that the proposed method is able to balance the performance tradeoff between sensing and computation functions, while reducing the energy consumption compared with the existing studies.
\end{abstract}


\section{Introduction}
\IEEEPARstart{F}{uture} 6G network will evolve into a multi-functional network that supports not only reliable data transmission from end to edge, but also ubiquitous intelligence applications with the features
of high-accuracy of sensing and low-latency of computation \cite{Feng2022MNET_Joint}.
Therefore, there is a surge of interest to explore the converging functionalities of sensing, communication, and computation, which is referred to as integrated sensing, communication, and computation (ISCC) \cite{XU2023WCL}.
Under ISCC networks, the users first perform radar sensing to obtain multi-view data, and then upload the sensed data to mobile edge computing (MEC) servers to enable the low-latency services. 

The performance of ISCC is typically hindered by unfavorable propagation conditions, particularly in disaster-stricken areas \cite{Duy2022JSAC}, remote areas, hot spots, and other scenarios with poor communication conditions \cite{Tran2022TWC}.
Unmanned aerial vehicle (UAV) has surfaced as a crucial enabling technology for boosting the capacity and wireless coverage owing to its superior ability of high mobility, full maneuverability, and low expense \cite{Wu2021_AComprehensive}. However, the high-mobility of UAVs may result in a dynamic network environment, thereby leading to increased complexity in facilitating lower energy consumption and real-time computation offloading performance.

Recent research reveals that DT is envisioned as an appealing technology for improving decision-making in optimizing service quality for time-varying wireless networks.
This paradigm can create a digital space model to evaluate the state information of entities in the physical networks, and allow for real-time monitoring of the network state \cite{Wu2021IOTJ_Digital,Alcaraz2022COMST_Digital}. In addition, more powerful AI technologies can be supported by DT to provide users with more timely decisions. Under this architecture, DT can replace the users and edge servers to make offloading decisions in the virtual space in advance, while the computing and
communication resources in the physical space can be provided quickly and accurately according to the request of users \cite{Wang2022IoT_Mobility,Duy2022LWC_Digital}. In this context, DT serves as a potential solution in 6G network to perceive the time-varying resource supply and demand, as well as achieve intelligent task scheduling and resource allocation, which is of paramount significance to the development of ISCC system.

To fully exploit the potential of employing UAV in ISCC networks, this paper presents the first attempt to introduce DT into ISCC networks to efficiently adjust the multidimensional network resources, and take full advantage of UAVs as edge servers by appropriately designing real-time UAV movement, thereby providing users with communication and computation services. However, the integration of heterogeneous network resources and dynamic information for real-time decision-making and long-term awareness imposes significant challenges in the research of UAV-aided computation offloading. As such, this paper proposes a multi-agent deep reinforcement learning (MADRL)-based scheme by considering the characteristics of distributive computation offloading, where DT is adopted to facilitate the centralized training and decentralized execution architecture. 
The main contributions of this work are summarized as follows.
\begin{enumerate}
    \item We propose a DT-empowered ISCC network by taking the cooperative relationship between the physical environment and the DT layer into account. Particularly, the users partially deliver their computational tasks to UAVs for edge processing and DT is leveraged to periodically estimate the practical computation requests of users and the operating states of UAV servers, where the mapping deviation of DT is considered.

    \item Different from the existing works that either optimize a single objective or a number of objectives via weighted sum, this paper aims to optimize the computation offloading energy consumption and the sensing beampattern gain simultaneously. To effectively address the challenging problem, we reformulate it as a Markov decision process (MDP) and apply the state-of-the-art multi-agent proximal policy optimization (MAPPO) method to capture the collaborative policy.
    
   \item To enhance the performance of training and accelerate the convergence speed, we apply Beta distribution and attention mechanism in actor and critic networks, respectively. Via numerical results, the rapid training convergence and effectiveness of our proposed scheme in optimizing the multi-objective problem are verified, while the superior performance of DT depends on the accuracy of DT estimation.
\end{enumerate}

The remainder of this paper is organized as follows. Section \ref{s:rw} reviews the related work. The system model is described in Section \ref{s:sys}. The multi-objective optimization problem is formulated in Section \ref{s:problem}. The proposed MAPPO algorithm for solving the formulated problem is presented in Section \ref{s:proposed}. In Section \ref{s:simulation}, the performance of our proposed algorithm is evaluated with detailed discussions. Finally, we conclude this paper in Section \ref{s:conclusion}.
 
\section{Related Work}\label{s:rw}
The applications of DT in MEC networks have gained growing attention to achieve real-time computing.
Specifically, \cite{Duy2022LWC_Digital} reflected the role of DT in MEC networks to minimize the end-to-end latency, where a joint optimization of transmit power, user association and task offloading is proposed.
The authors in \cite{Dai2021TII_Deep} considered a stochastic task arrival model in DT-enabled industrial applications and then applied the actor-critic-based DRL algorithm to minimize the long-term energy efficiency.
In \cite{Liu2022IoT_Digital}, a DT-assisted intelligent task offloading scheme was proposed and a value-based DRL method was leveraged to minimize the power and time overhead. 
In \cite{Zhou2022TII_Secure}, the authors proposed a DT-assisted algorithm to manage the resource scheduling and achieve long-term awareness.
In distributed networks, \cite{Zhang2022TII_Adaptive} utilized an MADRL algorithm to configure the task offloading and resource allocation in a DT-assisted MEC system via accommodating heterogeneous services.
Focusing on the flocking motion of UAVs, the authors in \cite{Shen2022IoT_Deep} used a DT-enabled MADRL framework to achieve higher average reward. However,
these works mainly focused on the centralized training at edge servers.

In the literature of ISCC networks, joint resource scheduling has been identified as a crucial factor in enhancing the performance of sensing, communication, and computation. For instance, 
an energy-efficient design for ISCC networks was proposed in \cite{Huang2022LWC_Integrated}, and the computational and communication resources were jointly optimized using an iterative algorithm with the assistance of intelligent reflecting surface. 
In \cite{Wang2022LCOMM_NOMA}, the authors proposed a non-orthogonal multiple access enabled integrated sensing-communication system, where the communication throughput and effective sensing power are jointly maximized. 
Aiming at maximizing overall performance and minimizing transmit power simultaneously, the authors of \cite{Qi2022TCOMM_Integrating} optimized the beamforming design in ISCC networks. As a step forward,
the authors of \cite{Ding2022JSAC_Joint} designed a multi-objective problem to minimize  the computational energy while optimizing the radar beampattern design in ISCC systems.
In \cite{Zhao2022TWC_Radio}, the authors investigated the wireless scheduling for ISCC with the aim of maximizing the throughput, while satisfying the heterogeneous requirements on resources.
The authors in \cite{Liu2022JSAC_Learning} proposed a deep learning-based approach to predict the beamforming matrix for the sum-rate maximization of vehicular networks.
These excellent achievements mainly utilize iterative algorithms to realize the resource allocation in the realm of ISCC.

Despite the aformentioned studies laid an initial foundation on the ISCC networks, they seldom consider the intelligent management of UAV-aided ISCC networks. In contrast to the above research, this paper aims to address this gap by focusing on the DT-empowered ISCC networks with multiple UAV edge servers. Specifically, a distributed training-based method with heterogeneous agents is designed to pursue the dynamically scheduling of the network resources as well as the configuration on sensing, communication, and computation.

\section{System Model}\label{s:sys}

\begin{figure*}[t]
	\centerline{\includegraphics[width=6.0 in]{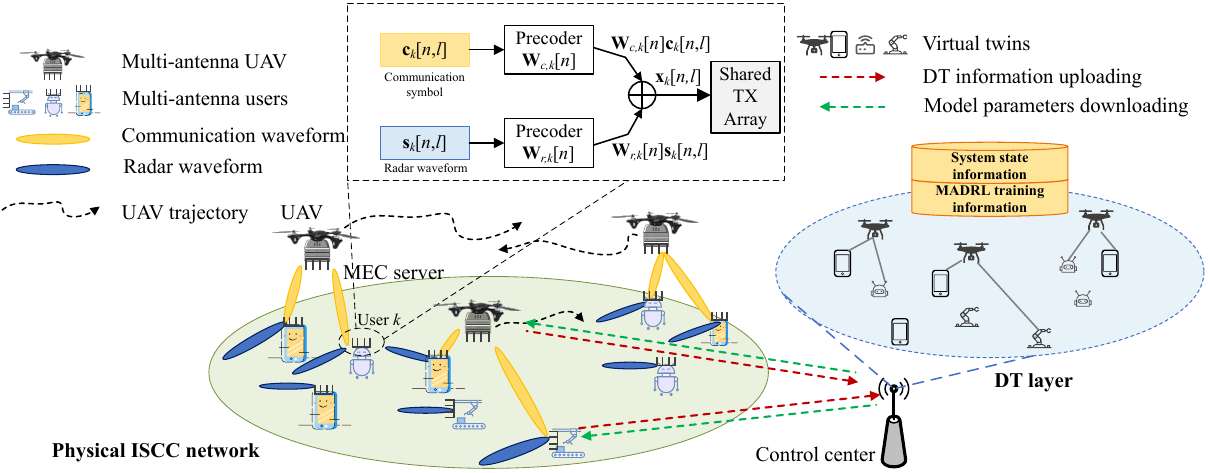}}
	\caption{System model of the DT-empowered ISCC networks.}
	\label{fig:sys-model}
\end{figure*}

We consider a DT-empowered ISCC network as shown in Fig. \ref{fig:sys-model}, which consists of $K$ users with $N_T$ antennas, $M$ UAVs with $N_R$ antennas, and a control center (e.g., BS).
Each user is equipped with dual function system and has wireless communication and radar detection functions at the same time. The user's radar can sense the surrounding environment and communicate with the UAVs to exchange control information and basic status. Meanwhile, the users will frequently generate computing-intensive tasks and the MEC servers are deployed at UAVs to accelerate the task processing through offloading. 
The DT layer is deployed at the control center to record the states of users and UAVs (e.g., channel information and service requirements) that facilitates the interaction between users and UAVs and then guides the edge computing service.
The offloading decision of each task is determined with the support of DT layer in terms of computing ability.
Note that a promising method to realize the user's DFRC is the transmitter shared by communication and radar sensing based on multi-beam, where the transmitted signal is the superposition of separately precoded communication symbols and radar waveforms. In addition, the same transmit antennas are shared by signals for sensing and communication.

To facilitate expression and analysis, the user set is defined as $\forall k \in \mathcal{K}\triangleq\{1,2,\cdots,K\}$ and the UAV set is $\forall m \in \mathcal{M}\triangleq\{1,2,\cdots,M\}$.
The UAVs have a flight period of $T$, which is divided into sufficiently short time slots with the length of $\delta_t 	= T/N$ such that the relative positions between UAVs and users are approximately unchanged in a given time slot but different in adjacent time slots.  
The set of time slots is recorded as $\forall n \in \mathcal{N}\triangleq\{1,2,\cdots,N\}$. 
We use the Cartesian coordinate system to simulate the positions of users and UAVs. Specifically, the time-varying horizontal position of UAV $m$ in time slot $n$ is ${\bf q}_m[n]=[x_m[n],y_m[n]]^{\rm T}$, the flying height is $H$ over the ground, and the position of ground user $k$ is ${\bf w}_k=[x_k,y_k]^{\rm T}$.
The displacement change of UAVs between different time slots is related to flight speed ${\bf v}_m[n]$ and acceleration ${\bf a}_m[n]$, and the collisions need to be avoided between UAVs, we thus have the following constraints
\begin{align}
	&{\bf q}_m[n+1]= {\bf q}_m[n] + {\bf v}_m[n] \delta_t +\frac{1}{2}{\bf a}_m[n] \delta_t^2,\\
	&\Vert {\bf q}_i[n] - {\bf q}_j [n] \Vert^2 \ge d_{\rm{min}}^2,
\end{align}
where $d_{\rm{min}}$ is the minimum safety distance between UAVs.

\subsection{Modeling of DT-empowered ISCC Network}
In this paper, the DT-empowered ISCC network consists of two types of entities, i.e., the users and the UAVs. 
To maintain the virtual twins, the users and the UAVs will upload the critical information of themselves to the DT layer at control center. Although DT model represents the operating state of the real network as accurately as possible, there are still mapping errors due to the limitations of the DT modeling method and the acquisition of modeling data. In addition, the information transmission randomness of wireless networks will further increase the mapping errors.

For each user $k$, the virtual twin needs to record its task information and location, which can be characterized by
\begin{align}
    {\rm DT}_k^u[n]=\{{\bf w}_k,\Omega_k[n],\tilde{f}_k[n]\},
\end{align}
where $\tilde{f}_k[n]$ denotes the estimated current computational resource for user $k$ to execute the task at time slot $n$, $\Omega_k[n]$ is the computational task information of users, which will be elaborated in \ref{subsec:comp}. 

For each UAV, the DT needs to reflect its scheduling of service, involving the allocation of resource and motivation status. Thus, the virtual twin of UAV $m$ can be characterized by
\begin{align}
    {\rm DT}_m^U[n]=\{{\bf q}_m[n],\alpha_{k,m},\tilde{f}_{k,m}[n],\forall k\in\mathcal{K}\},
\end{align}
where $\alpha_{k,m}$ and $\tilde{f}_{k,m}[n]$ are defined as the association factor of the network and the estimated computation resource allocated to user $k$ by UAV $m$, which will be illustrated in \ref{subsec:communication} and \ref{subsec:comp}.

The DT layer creates virtual twins of users and UAVs whose real-time states are synchronized with their counterparts in the physical world for further jointly optimization of heterogeneous resources.
Furthermore, DT performs an optimization framework to train MADRL models illustrated in Section \ref{s:proposed} and to download the decisions to the users and UAVs. Based on this fact, the network topology, the system state information, the virtual twins, and the MADRL training information are jointly managed by the DT layer.

\subsection{Communication Model}\label{subsec:communication}
For the communication function, 
user $k$ transmits $L$ symbols at time slot $n$ and the $l$-th symbol is given as\cite{Liu2020TSP_Joint}
\begin{equation}
    {\bf x}_k [n,l] = {\bf W}_{r,k}[n] {\bf s}_k[n,l] + {\bf W}_{c,k}[n] {\bf c}_k [n,l],
\end{equation}
where ${\bf s}_k[n,l]\in \mathbb{C}^{N_T \times 1}$ is an individual radar waveform, ${\bf W}_{r,k} [n] \in \mathbb{C}^{N_T \times N_T}$ denotes the precoding matrix of radar waveforms, ${\bf c}_k[n,l] \in \mathbb{C}^{d \times 1}$ represents $d$ communication symbols transmitted to UAVs, and ${\bf W}_{r,k} [n] \in \mathbb{C}^{N_T \times d}$ is the precoding matrix of communication symbols. 
According to \cite{Ding2022JSAC_Joint}, we note that the user's radar signal and communication signal are zero mean, time whitening, and generalized stationary random processes.
The user's radar waveform is not related to the communication signal. The covariance matrix of the same communication signal is unit matrix, and the covariance matrix between different communication signals is zero matrix. The covariance matrix of the same radar waveform is unit matrix, and the covariance matrix of different radar waveforms is zero matrix.

In the considered network, user $k$ uses the radar waveform for object detection, based on which the covariance of its transmission waveform is expressed as
\begin{align}
    \nonumber {\bf X}_k [n] &= \mathbb{E}\left[{\bf x}_k[n,l] {\bf x}_k[n,l]^{\rm H} \right] \\ 
     &= {\bf W}_{r,k}[n] {\bf W}_{r,k}^{\rm H} [n] + {\bf W}_{c,k}[n] {\bf W}_{c,k}^{\rm H}[n], 
\end{align}
and the transmission power of user $k$ is expressed as
\begin{equation}
    {\rm{tr}} \left( {\bf X}_k[n] \right) = {\rm{tr}} \left( {\bf W}_{r,k}[n] {\bf W}_{r,k}^{\rm H} [n] + {\bf W}_{c,k}[n] {\bf W}_{c,k}^{\rm H} [n]\right).
\end{equation}

In practical implementations, the channel between user $k$ and UAV $m$ is modeled as the Rician fading channel model, which is expressed as follows
\begin{small} 
\begin{equation}
    {\bf H}_{k,m}[n] = \sqrt{\frac{\psi_0}{d_{k,m}^2[n]} }\left( \sqrt{ \frac{\varsigma}{\varsigma +1} } {\bf \bar H}_{k,m}[n] +\sqrt{\frac{1}{\varsigma +1}} {\bf \hat H}_{k,m} [n] \right),
\end{equation}
\end{small}
where $\psi_0$ denotes the channel power gain at the reference distance, $d_{k,m}^2[n]=\Vert {\bf q}_m [n] - {\bf w}_k \Vert^2+H^2$, ${\bf \bar H}_{k,m}[n] \in \mathbb{C}^{N_R \times N_T}$ represents the line-of-sight channel component, ${\bf\hat H}_{k,m}[n] \in \mathbb{C}^{N_R \times N_T}$ represents the non-line-of-sight channel component. ${\bf\hat H}_{k,m}[n]$ follows the complex Gaussian distribution with 0-means and covariance matrix as the identity matrix, i.e., ${\bf\hat H}_{k,m}[n]\sim \mathcal{CN}\left( 0, {\bf I}_{N_R} \right)$, $\varsigma$ is the Rician factor specifying the power ratio. 

The signal received by UAV $m$ from user $k$ can be expressed as
\begin{small} 
\begin{align}
    \nonumber &{\bf y}_{k,m} [n]= \\ 
    &\alpha_{k,m}  {\bf H}_{k,m}[n] {\bf x}_k [n,l]\nonumber + \sum \limits_{i=1, i \ne k}^K{ \sum \limits_{j=1}^M {\alpha_{i,j} {\bf H}_{i,j}[n] {\bf x}_i[n,l]} }+ {\bf z}[n,l] \\
    =& \alpha_{k,m}  {\bf H}_{k,m}[n] {\bf W}_{c,k}[n] {\bf c}_k [n,l]\nonumber +\sum \limits_{i=1}^K{\sum \limits_{j=1}^M{\alpha_{i,j} {\bf H}_{i,j}[n] {\bf W}_{r,i}[n] {\bf s}_i[n,l]}} \\
    & +\sum \limits_{i=1,i \ne k}^K{\sum \limits_{j=1}^M{\alpha_{i,j} {\bf H}_{i,j}[n] {\bf W}_{c,i}[n] {\bf c}_i [n,l]}}+ {\bf z}[n,l],
\end{align}
\end{small}
where ${\bf z}[n,l]\sim \mathcal{CN} \left( 0,\sigma_c^2 {\bf I}_{N_k} \right)$ and $\sigma_c^2$ is the noise power.
The signal-to-interference-plus-noise ratio can be calculated as
\begin{align}
	{\bm \Gamma}_{k,m}[n]={\bf H}_{k,m}[n] {\bf W}_{c,k}[n] {\bf W}_{c,k}^{\rm H}[n]{\bf H}_{k,m}^{\rm H}[n] {\bf N}_{k,m}^{-1}[n],
	\label{subsec:commun1}
\end{align}
where
\begin{small} 
\begin{align}
    {\bf N}_{k,m}[n] \nonumber=& \sum \limits_{i=1}^K{\sum \limits_{j=1}^M{\alpha_{i,j} {\bf H}_{i,j}[n] {\bf W}_{r,i}[n] {\bf W}_{r,i}^{\rm H}[n] {\bf H}_{i,j}^{\rm H} [n]}}\\ \nonumber
   &+\sum \limits_{i=1,i \ne k}^K{\sum \limits_{j=1}^M {\alpha_{i,j} {\bf H}_{i,j}[n] {\bf W}_{c,i}[n] {\bf W}_{c,i}^{\rm H}[n] {\bf H}_{i,j}^{\rm H} [n]}} \\ 
   &+ \sigma_c^2 {\bf I}_{N_R}.
\end{align}
\end{small}

It follows from (\ref{subsec:commun1}) that
the transmission rate from user $k$ to UAV $m$ is given by
\begin{align}
	R_{k,m}[n]=B \log_2{\rm{det}}\left({\bf I}_{N_R} + {\bm \Gamma}_{k,m}[n]\right),
\end{align}
where $B$ is total available bandwidth. As a result, the transmission rate of user $k$ is given by
\begin{align}
    R_k[n] = \sum \limits_{m=1}^M{\alpha_{k,m}} R_{k,m}[n].
\end{align}

\subsection{Computation Model}\label{subsec:comp}
The wealth of sensing data is usually computation-intensive, user $k$ offloads its computation workload to UAV edge server to enable low-latency services. Defining a three tuple $\Omega_k[n]=(D_k[n],C_k[n],t_k^{\max}[n])$ at the beginning of each time slot, where $D_k[n]$ is the input data size of the generated computational task, and $C_k[n]$ is the average number of CPU cycles required to process unit bit of data in the task, and $t_k^{\max}[n]$  $(0\leq t_k^{\max}[n]\leq \delta_t)$ is the allowable maximum delay. We utilize the partial offloading mode and the task can be divided into two parts, where one part with the data size of $L_k^{o}[n]=\rho_k[n]D_k[n]$ offloaded to UAV for executing, and the other part with the data size of $L_k^l[n]=(1-\rho_k[n])D_k[n]$ calculated locally. $\rho_k[n]$ $(0\leq\rho_k[n]\leq 1)$ is defined as the task-partition factor. In addition, the two parts can be processed simultaneously. 

\subsubsection{Local computing}
It is understood that the DT layer can't fully represent the state of users and UAVs, especially for CPU frequency \cite{Liu2022IoT_Digital, Duy2022LWC_Digital}. 
We first express the estimated local computing time of user $k$ as
$    \tilde{t}_k^l[n] = \frac{ L_k^l[n] C_k [n] }{\tilde{f}_k^l[n]},
$
where $\tilde{f}_k^l[n]$ is the estimated value of user $k$'s CPU frequency. Hence, the gap of local computing time between DT estimation and actual value is given by   
\begin{equation} 
        \Delta t_k^l[n]=\frac{-L_k^l[n]C_k [n]{\hat{f}_k^l[n]}}{\tilde{f}_k^l[n](\tilde{f}_k^l[n]+{\hat{f}_k^l[n]})}.
\end{equation}
where $\hat{f}_k^l[n]$ is the estimated deviation of actual frequency $f_k^l[n]=\tilde{f}_k^l[n]+{\hat{f}_k^l[n]}$.

Accordingly, the actual value for local computing time is given by 
\begin{equation}   
t_k^l[n]=\tilde{t}_k^l[n]+\Delta t_k^l[n].
\end{equation}

\subsubsection{Computation offloading}
We define the association factor between user $k$ and UAV $m$ as $\alpha_{k,m}$. When user $k$ is associated with UAV $m$, we have $\alpha_{k,m} = 1$, otherwise $\alpha_{k,m} = 0$. 
For user $k$, the transmission delay in offloading is calculated by
$    t_k^o [n] = \frac{ \rho_k [n] D_k [n] }{ R_k [n] }.
$
Denoting the estimated value for allocated frequency of user $k$ by UAV $m$ as $\tilde{f}_{k,m}[n]$, the estimated computing time of UAV $m$ is given by
\begin{equation} 
    \tilde{t}_k^e [n] = \frac{ L_k^o[n] C_k[n] }{ \sum \limits_{m=1}^M{ \alpha_{k,m} [n] \tilde{f}_{k,m} [n] } }.
\end{equation}

Similar to the method above, the computing latency gap of UAV $m$ between DT and real value can be calculated as
\begin{equation} 
    \Delta t_{k,m}^e[n]=\frac{-L_k^o[n]C_k[n]{\hat{f}_{k,m}[n]}}{\tilde{f}_{k,m}[n](\tilde{f}_{k,m}[n]+{\hat{f}_{k,m}[n]})}.
\end{equation}
where $\hat{f}_{k,m}[n]$ is the estimated deviation of actual frequency $f_{k,m}[n]=\tilde{f}_{k,m}[n]+\hat{f}_{k,m}[n]$.

Hence, the actual value of edge computing time for user $k$ can be derived by
\begin{equation} 
    t_k^e[n]=\tilde{t}_k^e[n]+\sum\limits_{m=1}^M \alpha_{k,m}[n] \Delta T_{k,m}^{\text{comp}}[n].
\end{equation}
Based on the above discussions, the total latency imposed by computation offloading is calculated by $t_k^o[n] + t_k^e[n]$.
 
\subsection{Radar Sensing Model}
In this work, the radar receiver of each user can accurately obtain the transmitted communication symbol, and the communication signal can also be utilized for radar sensing. Therefore, the interference imposed by communication signal is negligible with respect to the radar receivers. During a duration for radar pulse repetition, the Doppler frequency shift caused by moving targets is usually assumed to be constant, so the range-Doppler parameters can be fully compensated \cite{LiuMIMO2018,Ding2022JSAC_Joint}. According to the radar target, if a far single point target is located at the $\theta_k$ direction, the echo received by user $k$ at time slot $n$ can be written as
\begin{align}
    {\bf y}_{k,r}[n,l]& = \psi_0{\bf A}_k ( \theta_k ) {\bf x}_k [n,l] + \sum \limits_{i=1, i \ne k}^K{{\bf H}_{k,i}  {\bf x}_i[n,l]} + {\bf z}_k[n,l].
\end{align}

Denoting ${\bf a}_{T,k}( \theta_k )  \in \mathbb{C}^{N_T \times 1}$ and ${\bf a}_{R,k}( \theta_k )  \in \mathbb{C}^{N_T \times 1}$ as the transmit and receive array steering vectors of the radar for user $k$, respectively, we have ${\bf A}_k ( \theta_k )={\bf a}_{R,k}( \theta_k ){\bf a}_{T,k}^{\rm H}( \theta_k )$. In addition,  $\omega_0$ denotes the Doppler frequency shift, ${\bf z}_k[n,l]$ denotes the additive white Gaussian noise with ${\bf z}_k[n,l] \sim \mathcal{CN} \left( 0, \sigma_R^2 {\bf I}_{N_T} \right)$, and ${\bf H}_{k,i} \in \mathbb{C}^{N_T \times N_T}$ is the channel interference from user $i$ to user $k$.
It is noteworthy that ${\bf a}_k ( \theta_k ) ={\bf a}_{T,k}( \theta_k )  = {\bf a}_{R,k}( \theta_k )  = \left[ 1, e^{j \frac{2 \pi}{\lambda} d_k \sin ( \theta_k )} , \cdots, e^{j \frac{2 \pi}{\lambda} d_k \left( N_T -1 \right) \sin ( \theta_k )} \right]^{\rm T}$, where $\lambda$ is the antenna spacing of users, $d_k$ is the signal wavelength, and we set $d_k=\lambda/2$.

In the case with multiple users, the signal interference between users imposed by radar affects the performance of radar detection. It is nature to consider the average interference-to-noise ratio (INR) as a constraint to guarantee the quality of the signal received from the radar sensing, based on which the average INR of user $k$ is given by

\begin{small} 
\begin{align}
    \nonumber\eta_k [n]& = \frac{ \mathbb{E}\left[ \sum \limits_{i=1,i \ne k}^K{ \Vert {\bf H}_{k,i} {\bf x}_i [n] \Vert_F^2 } \right]}{ \mathbb{E} \left[ \Vert {\bf z}_k  \Vert_F^2 \right] } \\ 
    &= \frac{ \sum \limits_{i=1,i \ne k}^K{\left( \Vert {\bf H}_{k,i} {\bf W}_{r,i}[n] \Vert_F^2 + \Vert {\bf H}_{k,i} {\bf W}_{c,i}[n] \Vert_F^2 \right)} }{N_T \sigma_R^2}.
\end{align}
\end{small}

\subsection{Energy Consumption Model}
\subsubsection{The energy consumption of users}
The effective capacitance coefficient of the CPU of user $k$ is $\kappa_1$. At the time slot $n$, the user $k$'s energy consumption during local computing is as follows
\begin{equation}
    E_k^l [n] =\kappa_1 f_k^l [n]^2 \left( 1-\rho_k[n] \right) D_k [n]C_k [n].
\end{equation}

The transmission energy consumption of user $k$ is expressed as
\begin{equation}    
	E_k^o[n] = t_k^o[n] \Vert {\bf W}_{c,k}[n] \Vert_F^2.
\end{equation}

According to the above analysis, the energy consumption of user $k$ yields
\begin{equation}    
	E_k [n] = E_k^l [n] + E_k^o [n].
\end{equation}

\subsubsection{The energy consumption of UAVs}
The effective capacitance coefficient of the CPU of the UAV is $\kappa_2$. When UAV $m$ provides computational service for user $k$ at time slot $n$, the computing energy is as follows
\begin{equation}
    E_m^e [n] = \sum\limits_{k=1}^K{ \kappa_2 \alpha_{k,m}f_{k,m}^2 [n] \rho_k [n] D_k [n] }.
\end{equation}

The flight power of UAV $m$ is calculated as
\begin{align}
	\nonumber p_m^{\rm{fly}}[n] = &P_0 \left( 1 + \frac{3 \Vert {\bf v}_m[n] \Vert^2} {U_{\rm{tip}}^2} \right) + \frac{1}{2} d_0 \rho s A \Vert {\bf v}_m[n] \Vert^3  \\ 
	&+ P_i \left( \sqrt {1 + \frac{\Vert {\bf v}_m[n] \Vert^4}{4 v_0^4}}  - \frac{\Vert {\bf v}_m[n] \Vert^2}{2v_0^2} \right)^\frac{1}{2},
\end{align}
where $P_0$ is the power of UAV's blade,  $P_i$ is the induced power during hovering, $v_0$ is the mean velocity of rotors. $U_{{\text{tip}}}$ is the blade's tip speed, $d_0$ is the fuselage drag ratio, $s$ is the rotor solidity, $A$ is the area of rotors, and $\rho$ denotes the air density.

Then, the flight energy consumption of UAV $m$ is calculated by
\begin{equation}	
	E_m^{\rm{fly}}[n]= p_m^{\rm{fly}}[n]\delta_t,
\end{equation}
and the energy consumption of UAV $m$ at time slot $n$ is expressed as
\begin{equation}	
E_m[n]= E_m^{\rm{fly}}[n]+E_m^e[n].
\end{equation}

\section{Problem Formulation}\label{s:problem}
The proposed ISCC network includes the functions of radar detection, computation offloading, and UAV trajectory planning. Therefore, the corresponding performance indicators and constraints should be clarified. 

\subsection{Radar Beampattern Design}
MIMO radar beampattern is an important design index of radar perception in the ISCC system, and high beampattern gain can be achieved in a given beam direction by carefully designing the covariance matrix of the sensing signal. Denoting the covariance matrix of the transmitted waveforms as ${\bf R}_{d,k}$, and the minimum square error problem is established as follows
\begin{subequations} \label{P0}
    \begin{align}
        \nonumber  \mathop{\min}\limits_{\mu_k,{\bf R}_{d,k}} &\sum \limits_{l=1}^L{\vert \mu_k P_{d,k}\left(\theta_l\right) - {\bf a}_k^{\rm H}\left( \theta_l \right) {\bf R}_{d,k}  {\bf a}_k \left( \theta_l \right) \vert^2}\\
        \text{s.t.}~
        &\mu_k  \ge 0, \forall k \in \mathcal{K},\\
        ~&{\rm{tr}} \left({\bf R}_{d,k} \right) = p_{\max},\forall k \in \mathcal{K}, n \in \mathcal{N},\\
        &{\bf R}_{d,k}  \succeq 0,{\bf R}_{d,k}  = {\bf R}_{d,k}^{\rm H}, \forall k \in \mathcal{K},
    \end{align}
\end{subequations}
where $p_{\max}$ is the maximum ISCC power of users, $P_{d,k}(\theta_l)$ is the ideal beampattern gain at angle $\theta_l \in [-\frac{\pi}{2},\frac{\pi}{2}]$, ${\bf a}_k(\theta_l)$ denotes the steering vector, $\mu_k$ is a scaling factor, and ${\bf R}_{d,k}$ is user $k$'s desired waveform covariance matrix.

\subsection{Multi-objective Optimization}
It is clear that the covariance matrix ${\bf R}_{d,k}$ can be designed via solving problem \eqref{P0}. However, the obtained ${\bf R}_{d,k}$ may not be suitable for actual radar design, due to the requirements of computational latency and the average INR of radar receiver. Similar to the previous work \cite{LiuMIMO2018,Ding2022JSAC_Joint}, we first minimize the constrained Frobeniusnorm square.
It is noteworthy that MIMO radars usually work with the maximum available power to enhance the sensing. Specifically, if the functions of ISCC networks focus on radar detection, it is better to reduce the communication transmission power, otherwise the radar sensing power has to be reduced.
Furthermore, the energy consumption of users and UAVs is a significant factor to evaluate the expense of the ISCC network, which can be represented by the weighted energy consumption.

Herein, we construct a multi-objective optimization problem (MOOP) by jointly designing the precoding matrix of radar waveform $\bm W_r\triangleq\{{\bf W}_{r,k}[n],\forall k\in \mathcal{K},n\in \mathcal{N}\}$, the precoding matrix of communication symbols $\bm W_c\triangleq\{{\bf W}_{c,k}[n],\forall k\in \mathcal{K},n\in \mathcal{N}\}$, the association factor of users $\bm \alpha \triangleq \{\alpha_{k,m},\forall k\in \mathcal{K},m \in \mathcal{M},n \in \mathcal{N}\}$, the CPU frequency of users $\tilde{\bm f}_i \triangleq \{\tilde{f}_k^l[n],\forall k\in \mathcal{K}, n \in \mathcal{N}\}$, the computational resource allocation of UAVs $\tilde{\bm f}_e \triangleq \{\tilde{f}_{k,m}[n],\forall k\in \mathcal{K}, m\in \mathcal{M}, n \in \mathcal{N}\}$, and the trajectory planning of UAVs $\bm q\triangleq\{{\bf q}_{m}[n],\forall m \in \mathcal{M},n \in \mathcal{N}\}$, which is in nature given as
\begin{subequations}\label{P12}
    \begin{align}
    &\nonumber  \mathop {\min} \limits_{\bm W_c,\bm W_r}  \sum \limits_{n=1}^N{ \sum \limits_{k=1}^K{ \Vert {\bf X}_k[n] - {\bf R}_{d,k} \Vert_F^2 } }, \\
    &\nonumber \min\limits_{\bm W_c,\bm W_r,\bm q, \tilde{\bm f}_i, \tilde{\bm f}_e, \bm \rho} \omega \sum\limits_{n=1}^N{ \sum\limits_{m=1}^M{E_m [n]} } + \sum\limits_{n=1}^N{\sum \limits_{k=1}^K{E_k [n]}}  \\
    \text{s.t.}~
        &\alpha_{k,m} \in \left\{ 0,1 \right\},\sum \limits_{m=1}^M{\alpha_{k,m}} \le 1,\forall k \in \mathcal{K},m\in\mathcal{M},\label{P1:alpha2}\\
        &{\rm{tr}}\left( {\bf X}_k[n] \right) = p_{\max} ,\forall k \in \mathcal{K},n \in \mathcal{N},\label{P1:p}\\
        &\eta_k[n] \le \zeta_k, \forall k \in \mathcal{K}, m \in \mathcal{M},\label{P1:inr}\\
        &{\rm{max}} \left\{ t_k^l [n], t_k^o [n]+ t_k^e [n] \right\} \le t_k^{\max}[n], \forall k \in \mathcal{K}, n \in \mathcal{N},\label{P1:delay}\\
        &0 \le \tilde{f}_k^l [n] \le f_k^{\rm{max}}, \forall k \in \mathcal{K}, n \in \mathcal{N},\label{P1:fk}\\
        &0 \le \tilde{f}_{k,m}[n] \le {f}_m^{\rm{max}},~0 \le \sum \limits_{k=1}^K{\alpha_{k,m} \tilde{f}_{k,m}[n]} \le f_m^{\rm{max}}, \label{P1:fmk}\\ 
        &0 \leq \rho_k[n]\leq 1,\forall k \in \mathcal{K}, n \in \mathcal{N}, \label{P1:rho}\\
        &\Vert {\bf a}_m[n] \Vert \le a_{\rm{max}},\Vert {\bf v}_m [n] \Vert \le v_{\rm{max}},\forall n \in \mathcal{N}, m \in \mathcal {M},\label{P1:v}\\
        &\Vert {\bf q}_i [n] - {\bf q}_j [n] \Vert^2 \ge d_{\rm{min}}^2,\forall i,j \in \mathcal{M}, i \ne j, \label{P1:q}
    \end{align}
\end{subequations}
where $\omega$ denotes the non-negative constant weight factor for UAV, $\zeta_k$ is the maximum tolerable INR level of user $k$, $f_k^{\max}$ is the maximum computational resources of user $k$, $f_m^{\max}$ is the maximum computational resources of UAV $m$, $a_{\rm{max}}$ is the maximum acceleration of UAV $m$, and $v_{\rm {max}}$ is the maximum flight speed of UAV $m$, and $\zeta_k$ is the maximum tolerable INR level of user $k$. Constraint \eqref{P1:alpha2} ensures that the user only associates to at most one UAV. Constraint \eqref{P1:p} specifies the transmission power of the user. Constraint \eqref{P1:inr} is the INR level required by user $k$. Constraint \eqref{P1:delay} indicates the tolerable computation delay, which is related to the DT estimation deviations $\tilde{f}_k^l [n]$ and $\tilde{f}_{k,m}[n]$. Constraint \eqref{P1:fk} limits the estimated computation resource $\tilde{f}_k^l [n]$ for user $k$ in DT layer. Constraint \eqref{P1:fmk} limits the estimated computation resource $\tilde{f}_{k,m}[n]$ allocated to user $k$ by UAV $m$  in DT layer. Constraint \eqref{P1:rho} is the task-partition factor.  Constraint \eqref{P1:v} are the acceleration and speed limitation of UAV. Constraint \eqref{P1:q} denotes the minimum safe distance between UAVs.

\section{DT-driven MADRL Approach}\label{s:proposed}
It can be readily derived that problem \eqref{P12} is a nonlinear and non-convex MOOP with highly-coupled and integer variables. This is very difficult to be solved by traditional offline optimization methods in the presence of time-varying channel conditions \cite{Wang2021TCCN_Multi,Peng2021JSAC_Multi,Zhao2022TWC_Multi}.
To tackle the challenge of addressing high-dimensional state and action spaces, in this section we consider to design an MAPPO-based training framework because it is capable of involving multiple types of policies to cooperatively and distributively decide the optimization variables.

\subsection{Modeling of Multi-agent MDP}
Since multiple UAVs and users participate in the network, the optimization problem has the characteristics of distribution in real scenarios. Therefore, our problem can be formulated as a multi-agent MDP. Typically, the elements of MDP involves a global state space $\mathcal{S}$, a global action space $\mathcal{A}$, and the reward function $\mathcal{R}$. In the multi-agent MDP, the state of environment is partially observable to agents, especially in privacy-awared systems and distributed frameworks. Denote the observation of agent $i\in\mathcal{I}\triangleq\{1,2,\ldots,I\}$ at time step $t$ as $o_t^i$, and thus the global state of environment $s_t$ can be obtained by combining the partial observations of agents.
To fully relieve the difficulty of decision-making for agents and pursue the near-optimal solutions, we consider to decompose the general policy on optimization variables into three policies. The global state space and action space can be respectively denoted as $\mathcal{S}=\mathcal{O}_1\times\ldots\times\mathcal{O}_I$ and $\mathcal{A}=\mathcal{A}_1\times\ldots\times\mathcal{A}_I$, which are extended as the Cartesian product of observation spaces $\mathcal{O}_i$ and action spaces $\mathcal{A}_i$ of all the agents. The three types of agents, which corresponding to three types of policies in the multi-agent system, are described as follows:

\subsubsection{Offloading-configuration agents}
This type of agents mainly focus on the offloading configuration for tasks. The index set of offloading-configuration agents is defined as $I_1\triangleq\{1,2,\ldots,K\}$. To decide the offloading proportion and association to UAVs, they need to observe their task information, locations of themselves and locations of UAVs.

\textbf{Observation}: The observation for offloading-configuration agents is as follows
\begin{equation}
    o_t^k=\left\{k, {\bf w}_k[n],{\bf q}_m[n], \Omega_k[n], \forall m \in \mathcal{M}\right\}.
\end{equation}

Note that each user can only obtain its own location via positioning service, and knows the information of all UAVs since the UAVs act as servers. For the huge difference on values of coordinates and task information, we scale them into $[0,1]$ according to the lower-bounds and upper-bounds of these variables decided by the ground width of region and the distribution of data size. In order to minimize the computational energy, the CPU frequency $\tilde{f}_k[n]$ can be simply set and estimated by following equation according to dynamic voltage frequency scaling technology \cite{Wang2016TCOMM_Mobile}
\begin{equation}  
 \tilde{f}_k[n]=\min\Big\{f_k^{\max},\frac{1}{t_k[n]} D_k[n]C_k[n]\Big\}.
\end{equation}

\textbf{Action}: The action for this type of agents should represent the decision variables, and thus can be defined as 
$
    a_t^k=\left\{ \rho_k[n], \alpha_{k,m}, \forall m \in \mathcal{M} \right\}.
$

For the constraint \eqref{P1:alpha2} on ${\bm \alpha}$, we select ${m_k}=\arg\max\limits_{m}\{\hat{\alpha}_{k,m}, \forall m \in\mathcal{M}\}$ as the associated UAV of user $k$, where $\hat{\alpha}_{k,m}$ is the output value of policy model. In addition, we let $\hat{\rho}_k[n]\leq 0$ to represent the fully local computing case for user $k$ at time slot $n$, where the range of output $\hat{\rho}_k[n]$ can be mapped into $[-\varepsilon,1]$ with $\varepsilon>0$. We then set $\rho_k[n]=\alpha_{k,m_k}=\lceil\hat{\rho}_k[n]\hat{\alpha}_{k,m_k}\rceil^+$ to ensure the feasibility of action, where $\lceil x\rceil^+=\max(0,x)$. 

\textbf{Reward}: The reward function of offloading-configuration agent needs to involve objective and penalty for not satisfying the latency requirements. The energy consumption of  users and their associated UAVs need to be decomposed for each user. Thus, the reward of agent $k$ is given by
\begin{equation}
    r_t^{k}=-\bar{E}_k^\omega[n]P_{T,k}^u(t),
\end{equation}
where 
\begin{equation}
    \begin{split}
         P_{t,T}^k=P\Bigg(&\sum\limits_{m=1}^M \alpha_{k,m} \max\left\{t_k^l[n],t_k^o[n]+t_k^e[n]\right\},\\
        &t_k^{\max}[n],t_k^{\max}[n]\Bigg),
    \end{split}
\end{equation}
and we calculate
\begin{equation}
    P(x,a,b)=2-\exp(-\lceil (x-a)/b \rceil^+)
\end{equation}
as the exponential penalty function, where
\begin{equation}
    \bar{E}_k^w[n]={\left( {\omega \sum\limits_{m=1}^M \alpha_{k,m} {E_m[n]}+ {E_k [n]}}\right)}
\end{equation}
denotes the average weighted energy consumption of user $k$. 

\subsubsection{Beampattern-configuration agents}
In the ISCC network, the users should balance the performance of sensing and communication by deciding the beampatterns for these functions. Denoting the index set of this type of agent as $\mathcal{I}_2\triangleq\{K+1,K+2,\ldots,2K\}$, we illustrate the MDP elements as follows:

\textbf{Observation}: The effect of beampatterns on objectives are associated with covariance matrix ${\bf R}_{d,k}$ and the relative position between user $k$ and their associated UAV. 
We define the observation of each beampattern-configuration agent as follows
\begin{equation}
    o_t^{K+k}=\{k,{\bf R}_{d,k}, {\bf H}_{k,m_k}[n],{\bf q}_{m_k}[n],{\bf w}_k[n],\Omega_k[n]\}.
\end{equation}

It is worth noting that deep neural networks are typically not able to address complex values.  For complex value $z=a+b{\rm j}$, we first transform it into $z=\vert z \vert {e}^{\rm j \angle z}$. In addition, $\vert z \vert$ and $\angle z$ are scaled into $[0,1]$. For instance, the entries of matrix ${\bf R}_{d,k}$ are rewritten into complex pairs $\{\vert z \vert, \angle z\}$, where the modular $\vert z \vert$ is divided by its trace $p_{\max}$. Furthermore, the modular of complex pairs for ${\bf H}_{k,m_k}[n]$ are normalized. Then, the pairs can be concatenated as vectors.

\textbf{Action}: To decide the beampattern for sensing and communication, the agents should give the action as follows
\begin{equation}
    a_t^{K+k}=\left\{ {\bf W}_{c,k}[n],{\bf W}_{r,k}[n]\right\}.
\end{equation}

To tackle constraint \eqref{P1:p}, the traces of matrices ${\bf W}_{c,k}[n]$ and ${\bf W}_{r,k}[n]$ can be given specifically by the output of neural networks and then be divided by ${\rm tr}({\bf X}[n])$.

\textbf{Reward}: Since the beampatterns ${\bf W}_{c,k}[n]$ and ${\bf W}_{r,k}[n]$ need to strike a balance between sensing and communication, the reward needs to consider the both functions. However, the part of objective function with respect to ${\bf W}_{c,k}[n]$ is much more sophisticated than that of ${\bf W}_{r,k}[n]$, and partial observation on channel state of each user sharply increases the difficulty for all agents to jointly provide optimal solution. To pursue a sub-optimal policy on beampatterns, we first evaluate the desired beampatterns for communication ${\bf R}_{d,k}^c[n]$, which have the maximum gain on the horizontal direction from users to the associated UAVs. The ${\bf R}_{d,k}^c[n]$ can also be added into input state-vector of neural networks. Therefore, the reward can be designed as 
\begin{align}
    \nonumber r_t^{K+k}= &\Big(2- \omega_s \varepsilon_s- (1-\omega_s)\varepsilon_c \Big)\cdot \\ &\Big(2-P_{t,T}^k+P(\eta_k[n],\zeta_k[n],\zeta_k[n])\Big)/2,
\end{align}
where
\begin{equation}
    \varepsilon_s=\frac{\Vert {\bf X}_k[n] - {\bf R}_{d,k} \Vert_F}{\Vert{\bf R}_{d,k} \Vert_F},
\end{equation}
    and 
\begin{equation}\varepsilon_c=\frac{\Vert {\bf W}_{c,k}[n] - {\bf R}_{d,k}^c \Vert_F}{\Vert{\bf R}_{d,k}^c \Vert_F}.
\end{equation}

\subsubsection{UAV agents}
The UAVs are required to control their speed, and allocate the CPU frequency for users. Denoting the index set of UAV agents as $\mathcal{I}_3\triangleq\{2K+1,2K+2,\ldots,2K+M\}$, the MDP elements are as follows:

\textbf{Observation}: Each UAV can obtain the location and task information of users served by it. Therefore, the observation can be expressed as 
\begin{equation}
    o_t^{2K+m}=\{ m, {\bf q}_m[n],{\bf q}_{-m}[n],{\bf w}_k[n], \Omega_k[n],\rho_k[n], \forall k \in \mathcal{K}_m \},
\end{equation}
where $\mathcal{K}_m$ denotes the users served by UAV $m$, and $-m$ denotes the indexes in set $\mathcal{M}\setminus \{m\}$.

\textbf{Action}: The UAVs need allocate their CPU frequency to execute users' tasks, and decide their movement to enhance the fairness of users. Therefore, the actions of users are given by
\begin{equation}
    a_t^{2K+m}=\{ {\bf a}_m[n], \tilde{f}_{k,m}[n], \forall k \in \mathcal{K}_m \}.
\end{equation}
 
Note that the output acceleration from policy model can be expressed as ${\bf \hat{a}}_m[n]=[\Vert{\bf a}_m[n]\Vert, \phi_m[n]]$, where $\phi_m[n]$ is the angular acceleration. The proportion of computational resource with respect to UAV's available CPU frequency and the proportion allocated for each user can also be represented by a vector with the length of $K+1$, where the entries of the users not served by UAV $m$ are multiplied by zero masks. Hence, the CPU frequency mapped from the action can be treated as an estimated value.

\textbf{Reward}: For UAV $k$, it needs to consider the weighted energy consumption as well as the distance to users in order to enhance the channel gain and fairness. Meanwhile, the collision and flying-out penalty should also be considered. Therefore, we design the reward as follows
\begin{align}
    \nonumber r_t^{2K+m}=&\Big(k_1\bar{E}_m^w[n]+ k_2P\big(\Vert {\bf q}_m[n]-\frac{1}{\vert\mathcal{K}_m\vert}\sum\limits_{k=1}^K \alpha_{k,m}{\bf w}_k[n]\Vert, \\
    &d_{\rm th},X\big)\Big) P_{t,T}^m P_{t,O}^m P_{t,C}^m ,
\end{align}
where $k_1$ and $k_2$ denote the adjusting factors, $d_{\rm th}$ is the threshold adjusting the distance from UAVs to users, and $X$ is the width of square service region. $\bar{E}_m^w[n]$ is the average energy consumption of UAVs, which is defined as
\begin{align} 
    \bar{E}_m^w[n]=\frac{1}{\vert\mathcal{K}_m\vert}  \sum\limits_{k\in\mathcal{K}_m} \alpha_{k,m} E_k[n] +  \varpi_m E_m[n].
\end{align}

The penalties are denoted by $P_{t,T}^m$, $P_{t,O}^m$, and $P_{t,C}^m$ respectively, where
\begin{align}
   \nonumber P_{t,T}^m =& \frac{1}{\vert \mathcal{K}_m \vert} \sum\limits_{k\in\mathcal{K}_m} P\Big( \alpha_{k,m} \max\{t_k^l[n],t_k^o[n]+t_k^e[n]\},    \\ 
    &t_k^{\max}[n],t_k^{\max}[n]\Big)
\end{align}
is the penalty for the latency requirements of users served by UAV $m$, 
\begin{align} 
    P_{t,O}^m =1+ \frac{1}{v_{\max}}\Vert{\bf q}_m[n]-{\rm clip}({\bf q}_m[n],0,X)\Vert
\end{align}
is the penalty when UAV tries to fly out of the square service region, and
\begin{align} 
    P_{t,C}^m =  \sum \limits_{j=1,j\neq m}^M P(d_{\min},\Vert {\bf q}_m[n] -{\bf q}_j[n] \Vert,d_{\min})
\end{align}
is the penalty for disobeying safety distance between UAVs.

\begin{figure*}[t]
    \centerline{\includegraphics[width=6 in]{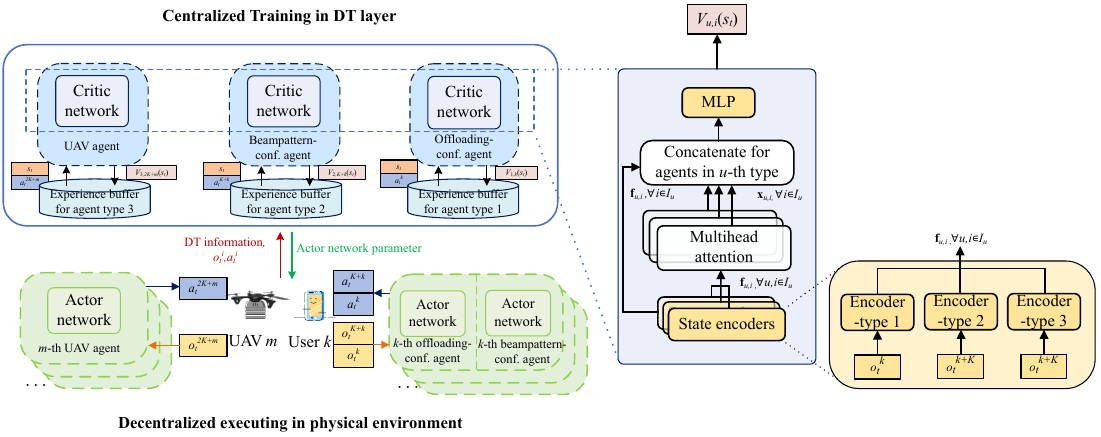}}
    \caption{The training framework of MADRL.}
    \label{fig:framework}
\end{figure*}

\subsection{DT for MADRL Training Framework}
It is widely acknowledged that MAPPO is an on-policy MADRL algorithm with state-of-art performance on various tasks \cite{Yu2021ar_Surprising}, which derives from trust policy optimization under the actor-critic framework. 
Accordingly, the continuous and discrete actions can be expressed by the output of actor network $\theta_u$, and the state-value function is evaluated by critic network $\omega_u$. The actor network for $u$-th type of agents represents the shared policy of the homogeneous agents, which is denoted as $\pi_{\theta_u}$. 

For the convenience of deployment in distributed DT-empowered ISCC networks, the centralized training and decentralized executing (CTDE) framework is utilized as shown in Fig. \ref{fig:framework}, where the agents execute actions with their actor networks, and train the centralized critic networks for each policy. Moreover, the rewards typically need the information of other agents, which are difficult to be independently evaluated by users and UAVs. To tackle this, we consider a CTDE framework enabled by DT, where the observations and actions are first collected during the interaction between agents and physical environment at each time step, and then are delivered to the DT layer for updating the virtual twins. Therefore, DT layer evaluates the rewards by the uploaded information, such as the actual computing time and the energy consumption. Meanwhile, the hyperparameters such as the distribution of task information can be timely estimated in DT layer. The global state $s_t$ is merged by observations in DT layer, and is sent to the critic networks of each type of agent, which work as the centralized state-value functions. 

During training process, the centralized state-value function of $u$-th type of agent is defined by
\begin{align}
V_{u,i}^{\pi}(s_t;\theta_u)=\mathbb{E}\left\{\sum\limits_{l=0}^\infty\gamma_u^l \mathcal{R}_{u,i}(s_{t+l},a_{t+l})|s_t=s,\pi \right\},
\end{align}
where $\mathbb{E}\{\cdot\}$ denotes the expectation operation, $a_t$ denotes the joint action of all the agents, $\mathcal{R}_{u,i}$ is the reward function for agent $i$ in $u$-th type of agents, and $\pi$ denotes the general policy of all agents. $\gamma_u$ denotes the discount factor that reveals the significance of future reward to agents. In addition, the action-value function is defined as
\begin{align}
    Q_{u,i}^{\pi}(s_t,a_t)=\mathbb{E}\left\{\sum\limits_{l=0}^\infty\gamma_u^l \mathcal{R}_{u,i}(s_{t+l},a_{t+l})|s_t=s,a_t=a,\pi \right\}.
\end{align}

As a result, we obtain the advantage function as 
\begin{align}
    A_{t,u,i}=Q_{u,i}^{\pi}(s_t,a_t)-V_{u,i}^{\pi}(s_t),
\end{align}
which can be estimated by state-value $V_u(s_t)$ as
\begin{align}
\hat{A}_u(s_t)= \sum\limits_{l=0}^{\infty}(\gamma_u\lambda)^l \Big(r_{t+l}+ \gamma_u V_u(s_{t+l+1})-V_u(s_{t}) \Big).
\end{align}

We utilize generalized advantage estimation (GAE) to evaluate the advantage function. The GAE factor $\lambda$ controls the tradeoff between variance and bias of reward, and $\delta_t= \left(r_{t}+ \gamma_u V_u(s_{t+1})-V_u(s_{t+l}) \right)$ is the temporal-difference error. 
Accordingly, the critic can be updated by following loss function
\begin{equation}\label{eq:L_critic}
    J(w_u)=\frac{1}{2}\left[ \hat{V}_{\omega_u}(s_t)-V_u(s_t) \right]^2.
\end{equation}

For the actor networks, PPO algorithm introduces a clipping factor $\epsilon$ to limit the update ratio of policy and thereby efficiently substitutes the calculation of trust region, based on which we can express the loss function of actor networks as 
\begin{align}
    \nonumber J(\theta_u)=&\mathbb{E}\Big\{\min\big[ \text{clip}(\frac{\pi_{\theta_u}(a_t|s_t)}{\pi_{\theta_u^{'}}(a_t|s_t)}, 1-\epsilon,1+\epsilon) \hat{A}_u(s_t), \\
    &\frac{\pi_{\theta_u}(a_t|s_t)}{\pi_{\theta_u^{'}}(a_t|s_t)} \hat{A}_u(s_t)
    \big]+ \psi S_{t,u} \Big\}, \label{eq:L_actor}
\end{align}
where $\theta_u^{'}$ represents the parameters of old policy for agent type $u$, $\frac{\pi_{\theta}(a_t|s_t)}{\pi_{\theta^{'}}(a_t|s_t)}$ is the update ratio, and $\psi S_{t,u}$ denotes the policy entropy representing the degree of exploration. In implement, the old policy $\theta_u^{'}$ can be  substituted by the log-probability of the actions and policy entropy $S_{t,u}$, which are stored in the experience buffers of DT layer after previous update. 
Therefore, the actor networks and critic networks can be updated by gradients $\nabla\theta_u=\frac{\partial J(\theta_u)}{\partial \theta_u}$ and $\nabla\omega_u=\frac{\partial J(\omega_u)}{\partial \omega_u}$. 

\subsection{New Attention Critic Mechanism}
\subsubsection{Beta policy}
It is worth noting that Gaussian distribution is widely adopted in the output of actor networks for policy-based DRL algorithms. However, in a majority of scenarios, the actions have both lower and upper bounds. On the contrary, Gaussian distribution is unbounded, and thus the action needs to be force clipped into given bounds, thereby leading to the boundary effects and estimation bias on policy gradient \cite{Chou2017ICML_Improving}. Furthermore, if the initial variance of Gaussian distribution is set small to reduce the boundary effect, the exploration ability will also decrease since the probability density will be more concentrated. If the variance becomes larger, the increasing probability of force-clipping on actions also makes the value of actions more likely to stay on boundaries, thereby leading to the reduction of exploration. Hence, we apply the Beta distribution for the output of actor networks. Denoting the parameters $\alpha$ and $\beta$, the Beta distribution with respect to $x$ is given by
\begin{align}
    f(x,\alpha,\beta)=\frac{\Gamma(\alpha+\beta)}{\Gamma(\alpha)\Gamma(\beta)}x^{\alpha-1}(1-x)^{\beta-1}. \label{eq:Beta}
\end{align}

It is obvious that \eqref{eq:Beta} has a bounded domain, thus being adaptive to the actions with double boundaries. In addition, it is beneficial for the algorithm to pursue a more uniform exploration at initial stage of training. Corresponding to this, the probability density of Beta distribution is typically higher closing to the boundaries than that of Gaussian distribution.

\subsubsection{Attention critic} The critic networks need to input large global state concatenated by observations. Therefore, the complexity of model sharply ascends when increasing the number of agents, which leads to the performance loss and slow convergence for typical full-connected networks. To address this difficulty, we utilize the attention mechanism in the critic networks. It is notable that attention mechanism can enhance the ability of agents to focus on the information of other agents in environment, some of which may have higher effect on the value function. It has been demonstrated that the learning speed and the performance can be scaled up by attention mechanism \cite{Cai2021TNSE_Cooperative}. The elements of attention mechanism is calculated as follows:

For agent type $u$, the observations $\{o_t^i,\forall i \in \mathcal{I}_u\}$ are first sent to the multi-layer perceptron (MLP) encoders, where the observations of different agent types are respectively encoded, to get the feature values $\{{\bf f}_{u,i},\forall i\in\mathcal{I}\}$. Subsequently, the feature values $\{{\bf f}_{u,i},\forall i\in\mathcal{I}\}$ are sent to attention heads of $u$-th type of agents, and thus the weighted attention values ${\bf x}_{u,i}$ are calculated by
$
{\bf x}_{u,i}=\sum\limits_{j\neq i}\alpha_{u,i,j} {\bf W}_{\rm val} {\bf f}_{u,j},
$
with
\begin{equation}
    \alpha_{u,i,j}=\text{Softmax}\left(\frac{{\bf f}_{u,j}^{\rm T} {\bf W}_{\rm key}^{\rm T}   {\bf W}_{\rm que} {\bf f}_{u,i} }{\sqrt{d_{\rm key}}}\right),
\end{equation}
where $\{{\bf f}_{u,j},\forall j\neq i\}$ are the feature values of other agents in $\mathcal{I}_u$ except $i$, $d_{\text{key}}$ denotes the variance of ${\bf x}_{u,i} W_{\rm que} W_{\text{key}}^\text{T}{\bf f}_{u,j}^\text{T}$. Matrix $W_{\rm que}$ transforms $f_{u,i}$ into a \textit{query}, matrix ${\bf W}_{\text{key}}$ transforms $f_{u,j}$ into a \textit{key}, and matrix $W_{\rm val}$ transforms $f_{u,j}$ into a \textit{value}.
Finally, ${\bf x}_{u,i}$ and $o_t^i$ are concatenated and then sent to the last MLP in the critic network of agent type $u$ to get the estimated state value $V_{\omega_u}(t)$.
Based on the above-mentioned discussions, we summarize the ATB-MAPPO training framework in Algorithm \ref{alg:mappo}.

\begin{algorithm}[t]
	\caption{Proposed ATB-MAPPO Training Framework}
    \label{alg:mappo}
    \begin{algorithmic}[1]
    \STATE{Initialize $n=1$, episode length $\rm{El}$, PPO epochs $\rm{Rp}$, and maximum training episodes ${\rm Me}$.}
    \STATE{Initialize actor networks $\theta_i$, critic networks $\omega_i$ on users and UAVs, $\forall e\in\{1,2,3\}$;}
    \FOR {Episode = $1,\dots,{\rm Me}$}
    \FOR {$t$ = $1,\dots,\rm{El}$}
            \IF{$n=1$}
            \STATE {Obtain beampattern ${\bf R}_{d,k}$ by solving problem \eqref{P0} $\forall \in \mathcal{K}$};
            \ENDIF
            \STATE {The agents of users obtain observations $o_t^i$ from environment, $\forall i \in \mathcal{I}_1\cup\mathcal{I}_2$;}
            \STATE {The agents of users execute actions $a_t^i$, $\forall i \in \mathcal{I}_1\cup\mathcal{I}_2 $;}
            \STATE {The agents of UAVs obtain $o_t^i$ from environment, $\forall i \in \mathcal{I}_3$;}
            \STATE {The agents of UAVs execute actions $a_t^i$, $\forall i \in \mathcal{I}_3$;}
            \STATE {Update $n = n$ mod $N$ + 1;}
            \IF{DT information uploading is required by DT layer}
            \STATE {The users and UAVs upload their observations and actions to DT layer;}
            \STATE{The DT layer updates the virtual twins and evaluates the rewards $r_t^i$;}
            \ENDIF
        \ENDFOR
        \STATE {Calculate log-probability $\text{pr}_t^i$ in the DT layer, $\forall i \in \mathcal{I},\forall t \in \{1,\ldots,{\rm El}\}$;}
        \STATE {The DT layer summarizes the transitions $\text{Tr}_t^i=$\{$o_t^i,a_t^i,r_t^i,s(t),\text{pr}_t^i$\}, $\forall i \in \mathcal{I},\forall t \in \{1,\ldots,{\rm El}\}$ in buffers;}
        \FOR {epoch = $1,\dots, {\rm Rp}$}
            \FOR {agents $i\in\mathcal{I}$}
                \STATE {Update actor and critic networks according to \eqref{eq:L_actor} and \eqref{eq:L_critic} by $\forall \text{Tr}_t^i \in \bm{B}_i$;}
            \ENDFOR
        \ENDFOR
    \ENDFOR
    \end{algorithmic}
\end{algorithm}

\section{Numerical Results}\label{s:simulation}
In this section, we carry out simulation experiments to evaluate the performance of proposed ATB-MAPPO training framework in DT-empowered ISCC network.  
We compare the performance of proposed scheme with the benchmarks as follows:
\begin{itemize}
    \item \textbf{Beta-MAPPO}: Proposed MAPPO-based training algorithm with Beta distribution on actor network and without attention mechanism.
    \item \textbf{Pure-MAPPO}: Proposed MAPPO-based training algorithm with widely adopted Gaussian distribution and without adopting attention mechanism \cite{Yu2021ar_Surprising}.
    \item \textbf{MADDPG}: The multi-agent deep deterministic policy gradient (MADDPG) algorithm, which is an off-policy MADRL algorithm with deterministic action output and noise for exploration \cite{Wang2021TCCN_Multi}. Each agent is corresponding to two shared actor and two critic networks. 
    \item \textbf{Random offloading}: The users randomly give actions, while the UAVs equally allocate the computational resource and move randomly.
\end{itemize}

\subsection{Simulation Scenario and Parameter Setting}
The simulation settings are illustrated as follows. We consider an ISCC network with a $1000$ m $\times$ $1000$ m square area. The UAVs and the users are uniformly located at the height of $200$ m and on the ground, respectively. The size of task data is uniformly distributed in [$0.5$ Mb, $D^{\max}$], where $D^{\max}$ is set to be $1.5$ Mb as default. The latency requirements of tasks $t_k^{\max}[n]$ is uniformly distributed in [$0.7$ s, $1.0$ s], and the average number of cycles required for each bit of task is $C_k[n]\in[500,1500]$ cycles. The estimation deviation of DT is set as 5\%, i.e., $|\tilde{f}_k[n]|\leq 5\%f_k[n]$. The path loss model for channels can be referred from \cite{Qi2022TCOMM_Integrating}. The average INR requirement is set as $\zeta_k = 0.7$ for all users. For algorithm setup, the value normalization is used and the reward is clipped into $[-5,5]$. The number of hidden layers for each MLP is $2$, and the sizes of hidden layers are set as $64$ and $128$ neurons. The length of feature values $V$ is $64$. 
Unless otherwise specified, the channel bandwidth is set as $B=10$ MHz, 
the noise power is $-65$ dBm, 
the Rician factor is $\varsigma=10$, 
the maximum ISCC power for users is set as $p_{\max}=0.5$ W, 
the number of antennas for users and UAVs is set as $N_T=N_R=4$, 
the channel power gain at the reference distance of $1$ m is $-30$ dB,
the capacitance coefficient is set as $\kappa_1=\kappa_2=10^{-27}$, 
the maximum CPU frequency of users and UAVs are set as $f_k^{\max}=1 $GHz and $f_m^{\max}=10$ GHz, respectively, 
the time period is set as $N= 40$ s, 
the duration of time slot $\delta_t= 1$ s, 
the maximum velocity of UAVs is $v_{\max}= 20 $ m/s, 
the maximum acceleration of UAVs $a_{\max}= 5 \text{m/s}^2$, 
the UAV settings $P_0,P_i,U_{\text{tip}},v_0,A$ and $s$ are set as $59.03$ W, $79.07$ W, $120$ m/s, $3.6$ m/s and $0.5030~\text{m}^2$ respectively, 
 the safe distance between UAVs is set as $d_{\min} =3$ m,
and the weight factor is $\omega=0.001$. 
For algorithm settings, we have the maximum training episodes $\text{Me}=300$ episodes, the episode length $\text{El}=200$ steps, the learning rate is $0.0005$, the discount factor is $\gamma_u=0.98$, the penalty factors $\mu_o$ and $\mu_t$ are the same as $0.1$, the number of PPO epochs $\text{Rp}=5$, the number of attention heads is $4$, the adjusting factors $k_1$ and $k_2$ are $0.3$ and $0.7$, respectively, the distance threshold $d_{\rm th}=350$ m, and the adopted optimizer is Adam.

\begin{figure}[t]
	\centering
	{\includegraphics[width=3.2in]{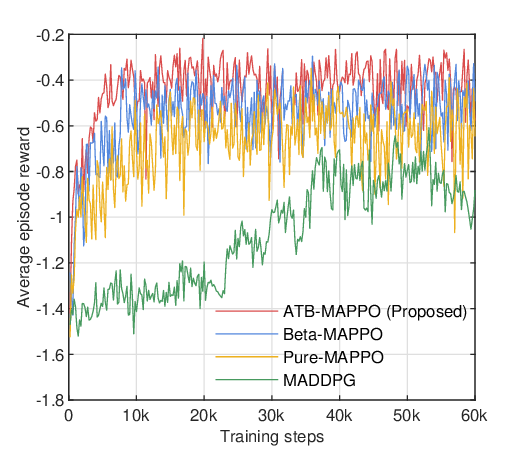}}
	\caption{Convergence versus offloading configuration agents.}
	\label{fig:rew-offload-alg}
\end{figure}

\begin{figure}[t]
	\centering
	{\includegraphics[width=3.2in]{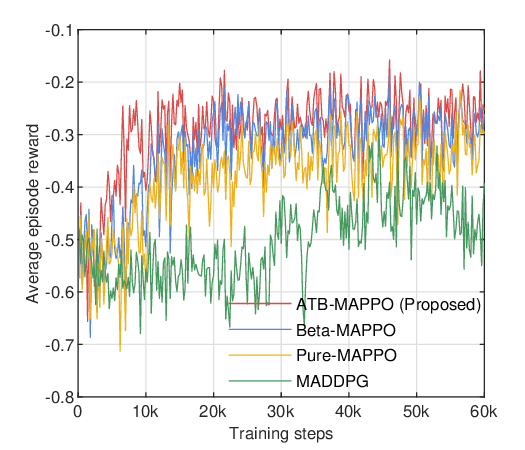}}
	\caption{Convergence versus UAV agents.}
	\label{fig:rew-UAV-alg}
\end{figure}

\begin{figure}[t]
	\centering
	{\includegraphics[width=3.2in]{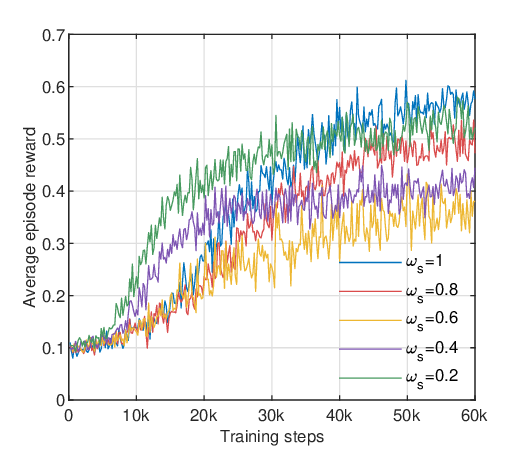}}
	\caption{Convergence versus different sensing-communication factors.}
	\label{fig:rew-beam-ws}
\end{figure}
\subsection{Performance Evaluation}
The convergence of deep neural network is extremely challenging and hard to be theoretically analyzed. The reason lies in that the convergence is highly dependent on DRL hyper-parameters, in which the quantitative relationship between deep neural network convergence and the hyper-parameters is sophisticated. Therefore, a reasonable choice of the hyper-parameters is required in order to achieve the convergence. We use numerical simulations (see Fig. 3-Fig. 5) to validate the convergence of the proposed PPO algorithm.

We first compare the convergence behavior of proposed ATB-MAPPO scheme with other MADRL benchmarks in Figs. \ref{fig:rew-offload-alg} and \ref{fig:rew-UAV-alg}, with $K=25$ users and $M=5$ UAVs. Intuitively, as the training steps increase, the reward of all the schemes gradually ascends, which confirms the effectiveness of MADRL algorithms in computation offloading. It is clear that the proposed ATB-MAPPO reaches at the highest reward, has a faster convergence rate than Beta-MAPPO, and has a higher reward than Pure-MAPPO with Gaussian distribution. This proves that Beta distribution is superior to Gaussian distribution in the actions of proposed ISCC network. As shown in Figs. \ref{fig:rew-offload-alg} and \ref{fig:rew-UAV-alg}, the convergence speed and the reward of MAPPO-based schemes achieve a remarkable improvement  compared to the MADDPG-based scheme.
As expected, it is readily found from Fig. \ref{fig:rew-offload-alg} that the reward of offloading-configuration agents gradually improves and the average episode reward of proposed ATB-MAPPO scheme achieves to about -0.40 as the highest value. 
Fig. \ref{fig:rew-UAV-alg} reveals that the UAV agents also improve their policy for trading off the energy consumption and relative position to users. Another observation is that the reward slightly declines at the initial stage of training, and this is mainly because the users are exploring to offload their tasks to UAVs, which is still learning to appropriately allocate the computational resource for users and control their speed, thereby leading to the growth of computational time for penalty and energy consumption.

We then evaluate the impact of sensing-communication weight factor $\omega_s$ on the reward of beampattern-configuration agents with $K=25$ users and $M=5$ UAVs. 
We can see from Fig. \ref{fig:rew-beam-ws} that as the $\omega_s$ varies from 0.2 to 0.6, the reward reduces in general but ascends during training. This result reveals that the agents are able to trade off the performance of two functions under the design of MSE-based reward. Furthermore, when $\omega_s>0.6$, the reward increases in general, where the radar beampattern has been paid more attention. We will further evaluate the effect of $\omega_s$ on beampatterns in the following Fig. \ref{fig:beam}.

\begin{figure}[t]
	\centering
	{\includegraphics[width=3.2in]{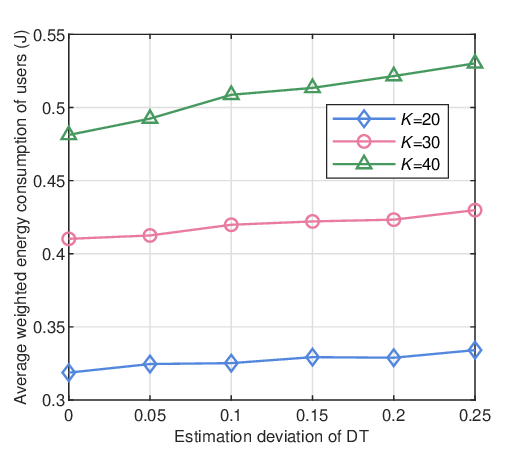}}
	\caption{The impact of estimation deviation of DT.}
	\label{fig:K-bias}
\end{figure}

\begin{figure}[t]
	\centering
	{\includegraphics[width=3.2in]{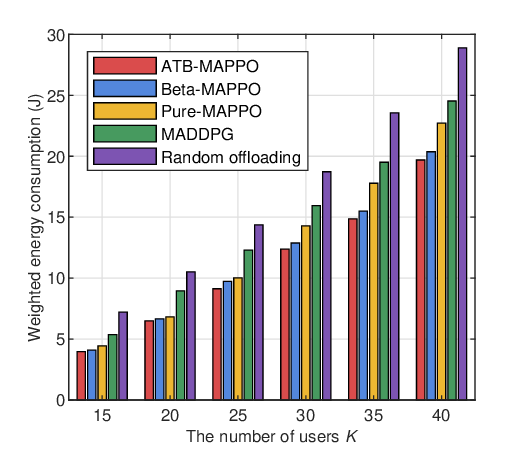}}
	\caption{Performance comparison versus different number of users.}
	\label{fig:K-alg}
\end{figure}

Fig. \ref{fig:K-bias} evaluates the impact of estimation deviation of DT on the system performance. Note that when the estimation deviation is 0, the perfect DT is achieved. It is observable that as the estimation deviation increases from 0\% to 25\%, the average weighted energy consumption of users has the tendency of increasing. It indicates that the deviation of CPU frequency may impose the fluctuating of computing time, the penalty, and thereby the quality of policy's distribution. Notably, the superior performance of DT comes from the accuracy of DT estimation.
Also, the result show that the  weighted energy consumption increases with an increasing $K$, where $K$ is chosen from
\{20, 30, 40\}.

Fig. \ref{fig:K-alg} compares the weighted energy consumption under different number of users with $M=5$ UAVs. It can be readily observed that as the number of users becomes large, the weighted energy consumption of UAVs and users accordingly grows. As expected, the proposed ATB-MAPPO has the best performance, and the performance gaps between MADDPG and MAPPO-based schemes remain large. Another observation is that the performance gap between adjacent settings has the trend of increasing, which means that the average weighted energy of users also increases. This is because as more users join in the network, the signal interference between users grows and the transmission rate reduces, resulting in more computational resource is required for less available computing time with respect to latency requirements.

\begin{figure}[t]
	\centering
	{\includegraphics[width=3.2in]{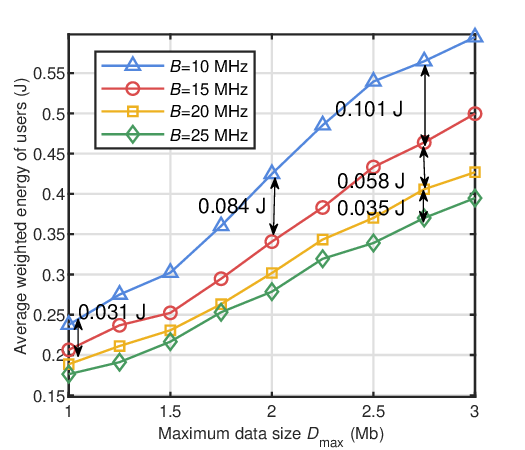}}
	\caption{The performance versus different task size and bandwidth.}
	\label{fig:D-B}
\end{figure}

\begin{figure}[t]
	\centering
	{\includegraphics[width=3.2in]{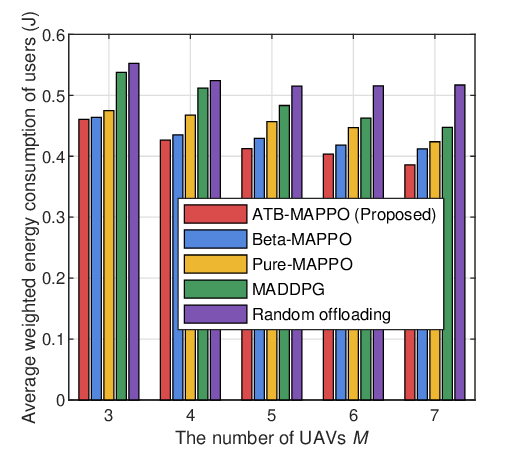}}
	\caption{Performance comparison versus different number of UAVs.}
	\label{fig:UAV-alg}
\end{figure}

Fig. \ref{fig:D-B} evaluates the average weighted energy of users under different settings of bandwidth and task size with $K=20$. We can find that as the maximum task size $D^{\max}$ grows, the energy consumption of users gradually grows, while the energy consumption reduces as the bandwidth grows. This can be explained by the fact that the increase of communication resource enhances the transmission rate of users, and the increase of task size makes the average computational energy increases. It can also be seen that the performance gap between different bandwidth gradually increases. This mainly because the increase of bandwidth makes the users willing to offload more tasks, which relieves the local computational load. In addition, less transmission time spares the available time for computation. Therefore, the CPU frequencies of users and UAVs can be jointly saved, leading to the rapid decreases of computational energy.

\begin{figure}[t]
	\centering
	\begin{minipage}[t]{0.50\textwidth}
		\centerline{\includegraphics[width=\textwidth]{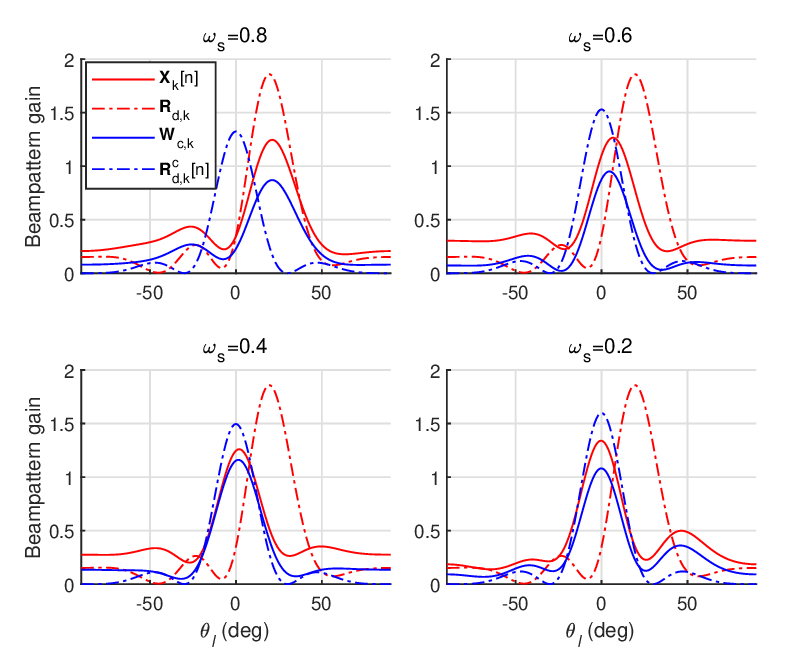}}
		\caption{The examples of beampatterns for sensing and communication.}
		\label{fig:beam}
	\end{minipage}
\end{figure}

In Fig. \ref{fig:UAV-alg}, we compare the performance obtained by the five schemes versus different number of UAVs under $K=30$ users. It can be  observed that as the number of UAV grows, the average weighted energy consumption of users reduces, verifying the fact that more UAVs provide more computational resource, and the agents can seek the balance between computational load on UAVs and users to reduce the energy consumption. Furthermore, the gap between MADDPG and MAPPO-based schemes gradually reduces as the number of UAVs increases, and the proposed algorithm constantly outperforms benchmarks. As such, we have presented that the proposed ATB-MAPPO scheme can efficiently optimize the policy for weighted energy consumption minimization in the service.

We then present the examples of optimized beampattern gain from $-90\deg$ to $90\deg$ with respect to sensing-communication weight factor $\omega_s$ in Fig. \ref{fig:beam}. The azimuth angle for the evaluated user to its associated UAV is set to be 0, the beampattern gain for covariance matrix ${\bf A}$ at $\theta_l$ is calculated as ${\bf a}(\theta_l)^{\rm H}{\bf A}{\bf a}(\theta_l)$, where ${\bf a}(\theta_l)$ is the steering vector at $\theta_l$. We can see that when $\omega_s=0.2$ and $\omega_s=0.4$, the primary beams of ${\bf X}_{k}[n]$ and ${\bf W}_{c,k}[n]$ focus on the direction for communication. When $\omega_s$ becomes larger, the primary beams move to fit better with the desired beampattern ${\bf R}_{d,k}$ of sensing.  Hence, the performance for sensing and communication can be approximately traded off by $\omega_s$. In this regard, the effectiveness of reinforcement learning method for tackling beampatterns and channels is verified.

\begin{figure}[t]
	\centering
	{\includegraphics[width=3.2in]{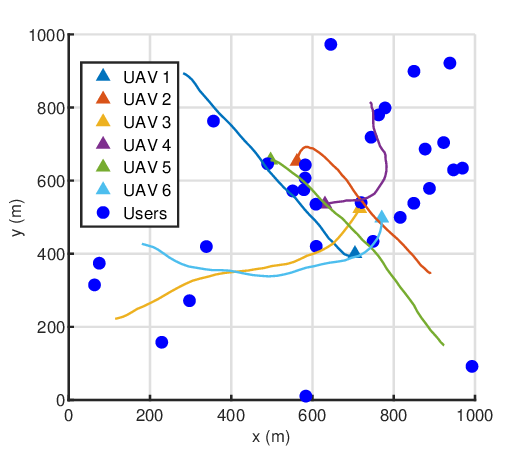}}
	\caption{The example of trajectories of UAVs under $K=30$ and $M=6$.}
	\label{fig:traj2}
\end{figure}

In Fig. \ref{fig:traj2}, we present the example of UAVs' trajectories  with $K=30$ and $M=6$. It can be observed that UAVs are able to select the regions with
more users, and can adaptively update their positions according to the distribution of associated users. Moreover, it implies that the reward can guide the UAVs to find the relative fair area for users and then hover slowly to save flying energy. The trajectories are jointly smooth, which makes it applicable for practical UAV movement compared with simply controlling the direction and velocity like \cite{Wang2021TCCN_Multi}, revealing the effectiveness of UAV agents in trajectories design.

\section{Conclusion}\label{s:conclusion}
In this paper, we proposed a multi-UAV-enabled ISCC network to jointly optimize the computation offloading and sensing performance. A cooperative MADRL framework was developed to solve the challenging multi-objective optimization problem. Considering the high-dimensional hybrid action spaces, we introduced the MAPPO method with attention mechanism and Beta distribution to obtain the optimal learning strategy effectively. Numerical results verified that the proposed scheme can significantly reduce the network energy consumption compared with the benchmark approaches. Meanwhile, the obtained radar beampattern can fit the ideal radar beam well. 

\bibliographystyle{IEEEtran}
\bibliography{IEEEabrv,refs}

\end{document}